\documentclass[journal]{vgtc}                     % final (journal style)
\onlineid{1544}

\vgtccategory{Research}

\vgtcpapertype{Theoretical \& Empirical}

\title{How Aligned are Human Chart Takeaways and LLM Predictions?\\ A Case Study on Bar Charts with Varying Layouts}

\author{%
  Huichen Will Wang,
  Jane Hoffswell, 
  Sao Myat Thazin Thane,
  Victor S.\ Bursztyn, and
  Cindy Xiong Bearfield
}

\authorfooter{
  \item
  	Huichen Will Wang is with the University of Washington.
  \item
  	Jane Hoffswell and Victor S.\ Bursztyn are with Adobe Research.
  \item Sao Myat Thazin Thane is with UMass Amherst.
  \item Cindy Xiong Bearfield is with the Georgia Institute of Technology.
}

\abstract{%
Large Language Models (LLMs) have been adopted for a variety of visualizations tasks, but how far are we from perceptually aware LLMs that can predict human takeaways? Graphical perception literature has shown that human chart takeaways are sensitive to visualization design choices, such as spatial layouts. In this work, we examine the extent to which LLMs exhibit such sensitivity when generating takeaways, using bar charts with varying spatial layouts as a case study. We conducted three experiments and tested four common bar chart layouts: vertically juxtaposed, horizontally juxtaposed, overlaid, and stacked. 
In Experiment 1, we identified the optimal configurations to generate meaningful chart takeaways by testing four LLMs, two temperature settings, nine chart specifications, and two prompting strategies. We found that even state-of-the-art LLMs struggle to generate semantically diverse and factually accurate takeaways. 
In Experiment 2, we used the optimal configurations to generate 30 chart takeaways each for eight visualizations across four layouts and two datasets in both zero-shot and one-shot settings. Compared to human takeaways, we found that the takeaways LLMs generate often do not match the types of comparisons made by humans. In Experiment 3, we examined the effect of chart context and data on LLM takeaways.  We found that LLMs, unlike humans, exhibit variation in takeaway comparison types for different bar charts using the same bar layout. 
Overall, our case study evaluates the ability of LLMs to emulate human interpretations of data and points to challenges and opportunities in using LLMs to predict human chart takeaways.
}


\teaser{
  \centering
  \includegraphics[width = 15cm]{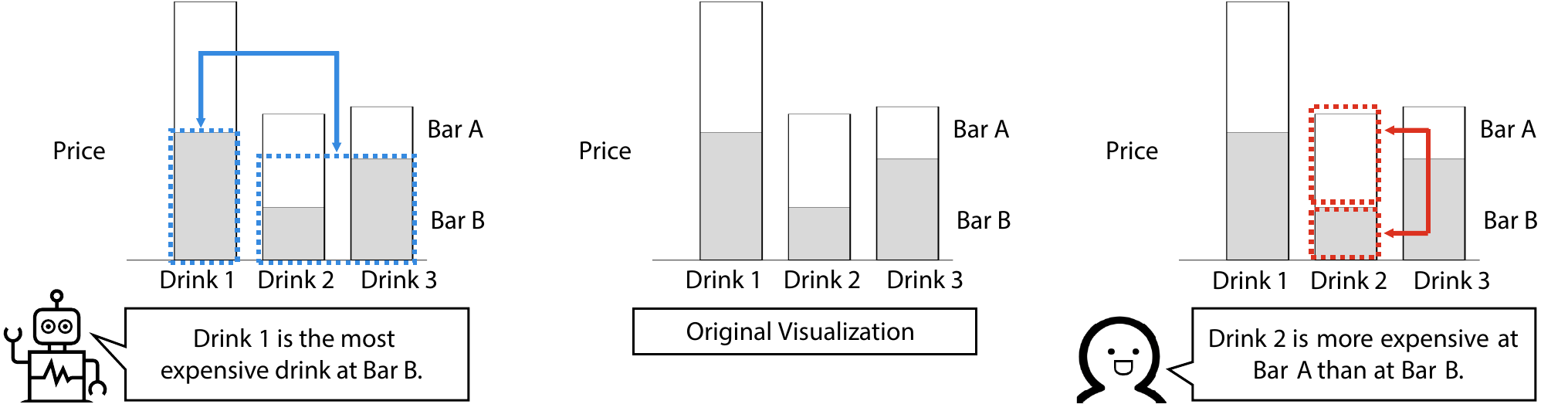}
  
 \caption{There is a discrepancy between human chart takeaways and predictions of human chart takeaways generated by large language models. For a chart that shows the price of three drinks in two bars, a human would tend to compare the price of Drink 2 between the two bars, but the model predicts a human to compare the price of the three drinks in Bar B.}
   \label{fig:teaser}
}

\graphicspath{{figs/}{figures/}{pictures/}{images/}{./}} % where to search for the images

\usepackage{tabu}                      % only used for the table example
\usepackage{booktabs}                  % only used for the table example
\usepackage{lipsum}                    % used to generate placeholder text
\usepackage{mwe}                       % used to generate placeholder figures
\usepackage{tabularx}
\usepackage{listings}
\usepackage{mdframed}
\usepackage{amsmath}

\usepackage{mathptmx}                  % use matching math font

\usepackage{xargs}      % Used for new commands with optional arguments
\usepackage{soul}       % Used for custom comments
\usepackage{color}      % Used for custom colors in comments
\usepackage{xspace}     % Used for abbreviation spacing
\usepackage{xpunctuate} % Used for abbreviation spacing
\usepackage{multirow}
\usepackage{array}
\usepackage{multirow}
\usepackage{tabularx}
\usepackage{listings}
\usepackage{mdframed}
\usepackage{pifont}     % Used as symbols for small table

\newcommand{\ie}{{i.e.,}\xspace}
\newcommand{\eg}{{e.g.,}\xspace}
\newcommand{\etal}{{et~al\xperiod}\xspace}

\newcommand{\cmark}{\ding{51}}
\newcommand{\xmark}{\ding{55}}

\definecolor{lightpink}{RGB}{237,157,202}
\definecolor{lightred}{RGB}{210,121,121}
\definecolor{lightorange}{RGB}{230,170,50}
\definecolor{lightgold}{RGB}{210,194,121}
\definecolor{lightgreen}{RGB}{121,210,121}
\definecolor{lightaqua}{RGB}{121,206,210}
\definecolor{lightblue}{RGB}{121,124,210}
\definecolor{lightpurple}{RGB}{153,102,255}
\definecolor{red}{RGB}{178,34,34}
\definecolor{gray}{RGB}{166,166,166}
\definecolor{forestgreen}{RGB}{74,103,65}

\newcommand{\gray}[1]{\textcolor{gray}{#1}}

\definecolor{CindySalmon}{RGB}{232, 125, 114}

\newcommandx{\guest}[3][1=]
    {\setulcolor{lightorange}{\ul{#1}} \textcolor{lightorange} %% Usage: \guest[Underline]{Name}{Comment}
    {[\textbf{#2:} #3]}}
\newcommandx{\cx}[2][1=] 
    {\setulcolor{CindySalmon}{\ul{#1}} \textcolor{CindySalmon}
    {[\textbf{Cindy:} #2]}}
\newcommandx{\victor}[2][1=] 
    {\setulcolor{lightgold}{\ul{#1}} \textcolor{lightgold}
    {[\textbf{Victor:} #2]}}
\newcommandx{\jane}[2][1=] 
    {\setulcolor{lightgreen}{\ul{#1}} \textcolor{lightgreen}   %% Usage: \jane[(optionally) underline text]{With a comment.}
    {[\textbf{Jane:} #2]}}    
\newcommandx{\will}[2][1=] 
    {\setulcolor{lighblue}{\ul{#1}} \textcolor{lightblue}
    {[\textbf{Will:} #2]}}
\newcommand{\add}[1]{{#1}}
    
\newcommand{\feedbackProvided}{}

\newcommand{\figureRevenue}{
\begin{figure}[t]
    \centering
    \includegraphics[width=\columnwidth]{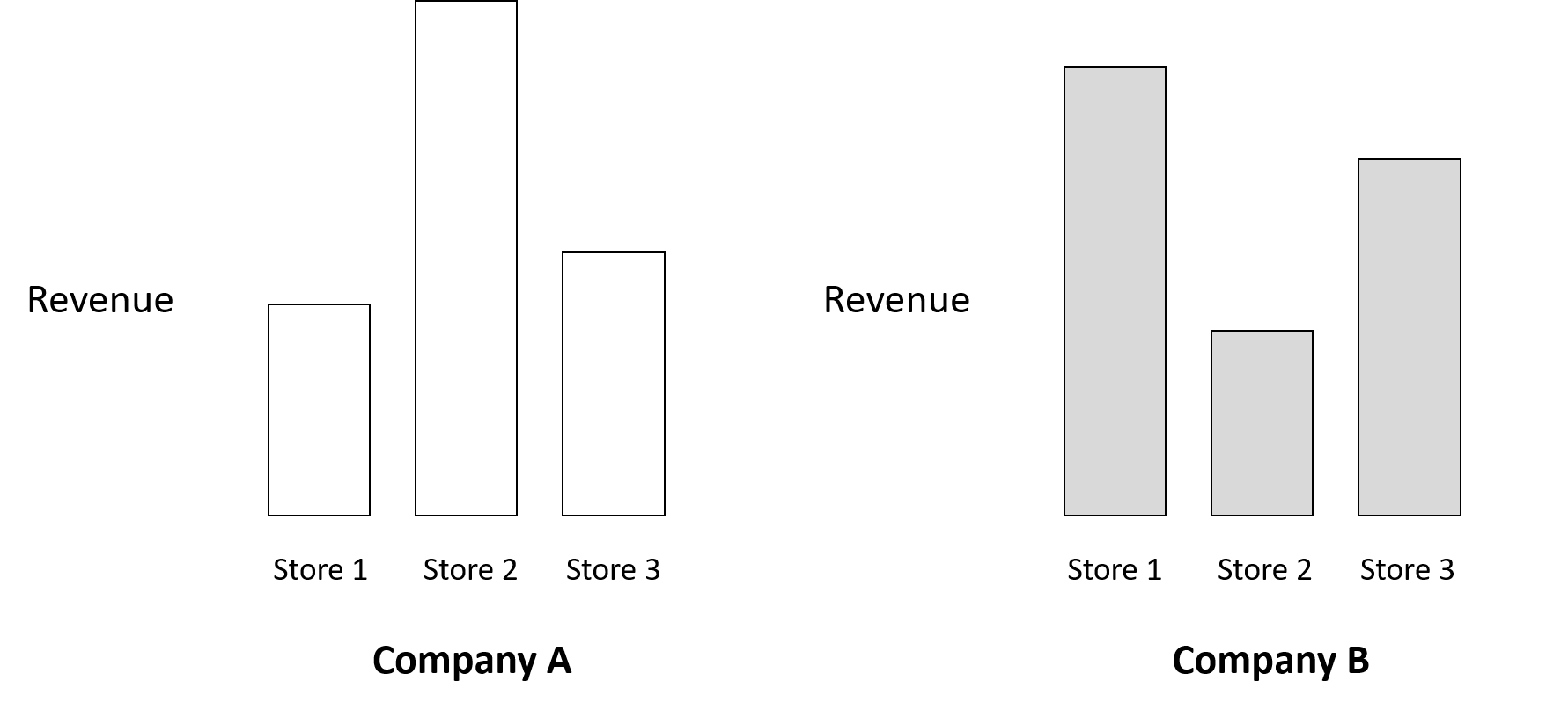}
    \caption{A horizontally juxtaposed bar chart depicting the revenue of three stores from two companies. Figure is from Xiong~\etal~\cite{xiong2021visual}.}
    \vspace{-3mm}
    \label{fig:revenue}
\end{figure}
}

\newcommand{\summary}{
\begin{figure*}[t]
    \centering
    \includegraphics[width = \linewidth]{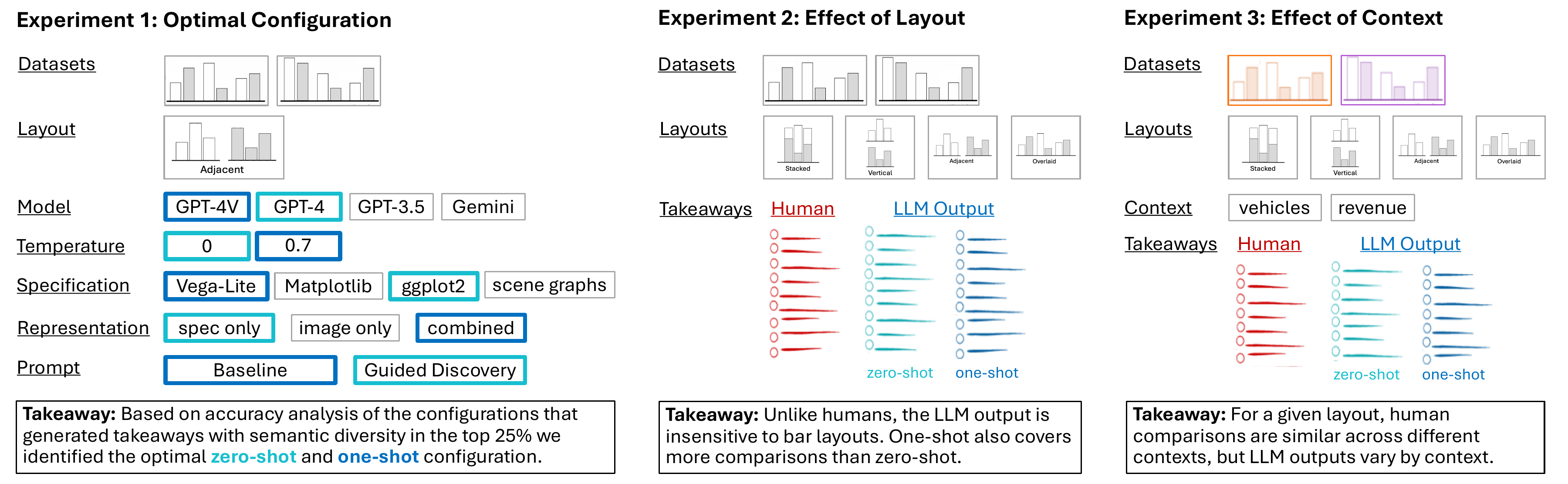}
    \caption{Our case study includes three experiments. In Experiment 1, we varied the LLM, decoding temperature, chart specification, and prompting strategy, and identified optimal configurations to elicit LLM chart takeaways for both zero-shot and one-shot settings. In Experiment 2, we generated takeaways using optimal configurations and examined whether LLMs’ comparisons are perceptually sensitive to bar arrangement like humans are. In Experiment 3, we examined whether LLMs’ comparisons are insensitive to data and context like humans are.}
         \label{fig:summary}
    \vspace{-3mm}
\end{figure*}
}

\newcommand{\clusters}{
    \begin{figure}[t]
        \centering
        \includegraphics[width=\columnwidth]{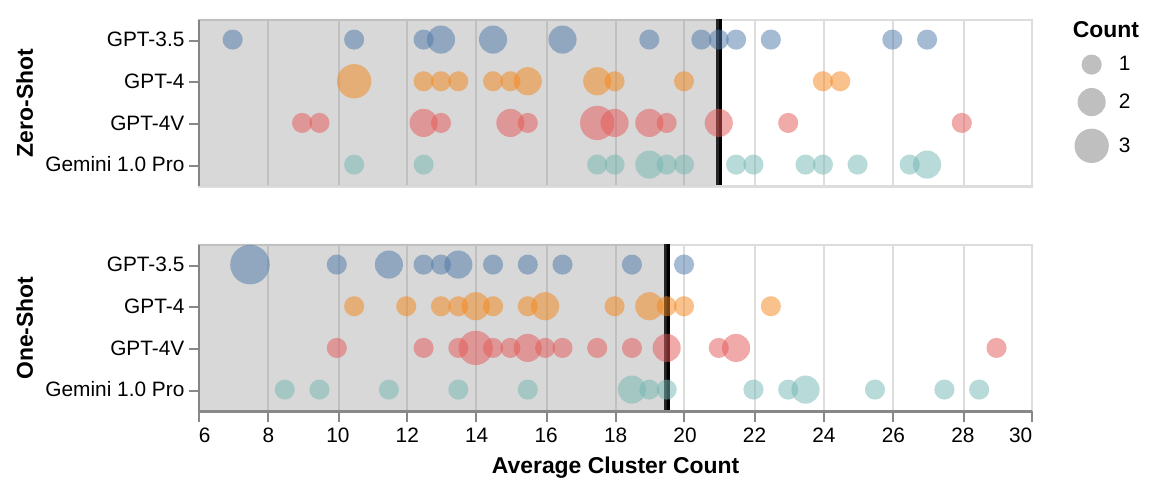}
        \caption{\add{Distribution of the average cluster count for each configuration (broken down by the LLM type) for the zero-shot and one-shot setting. We reviewed the accuracy of the top 25\% (see Section~\ref{semantic diversity}), corresponding to thresholds of 21 and 19.5 for the zero-shot and one-shot settings.}}
        \label{fig:cluster counts}
    \end{figure}
}

\newcommand{\llmcontextdata}{
    \begin{figure*}[h!]
        \includegraphics[width=0.95\textwidth]{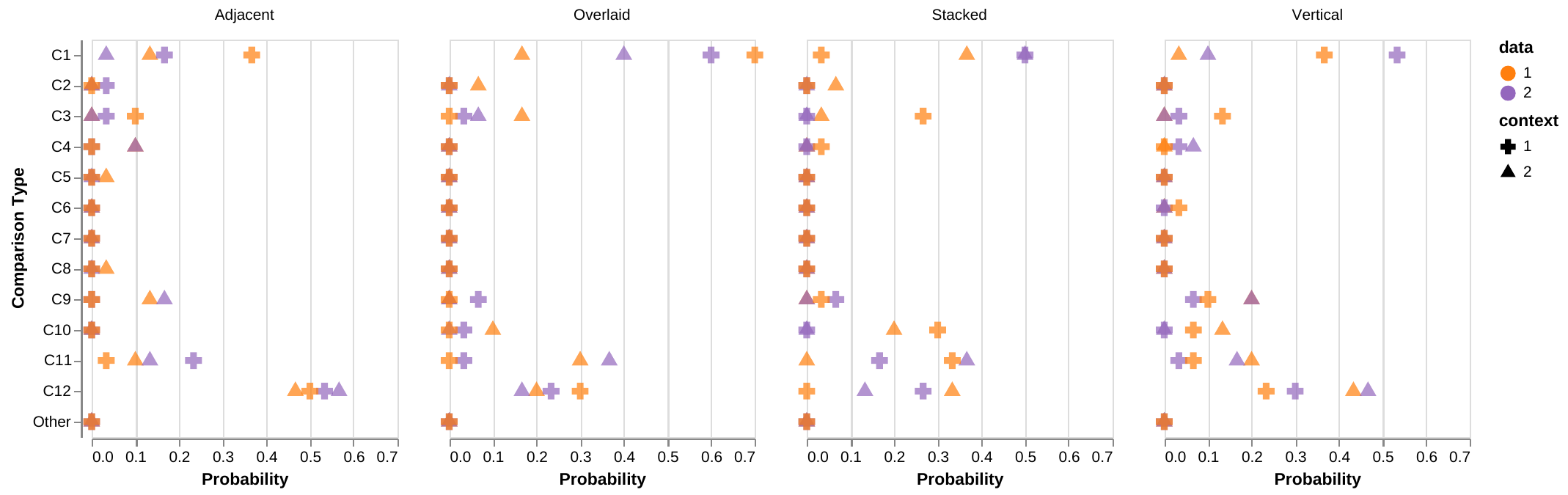}
        \caption{Distributions of LLM comparison types for the four examples (2 contexts $\times$ 2 datasets) for each layout. With the exception of the stacked layout, differing contexts (same color, different shapes) induce greater variations in the distributions than differing data (same shape, different colors).}
        
        \label{fig:llmcontextdata}
        \vspace{-3mm}
    \end{figure*}
}

\newcommand{\humanllmexample}{
    \begin{figure*}[h!]
        \centering
        \includegraphics[width=\textwidth]{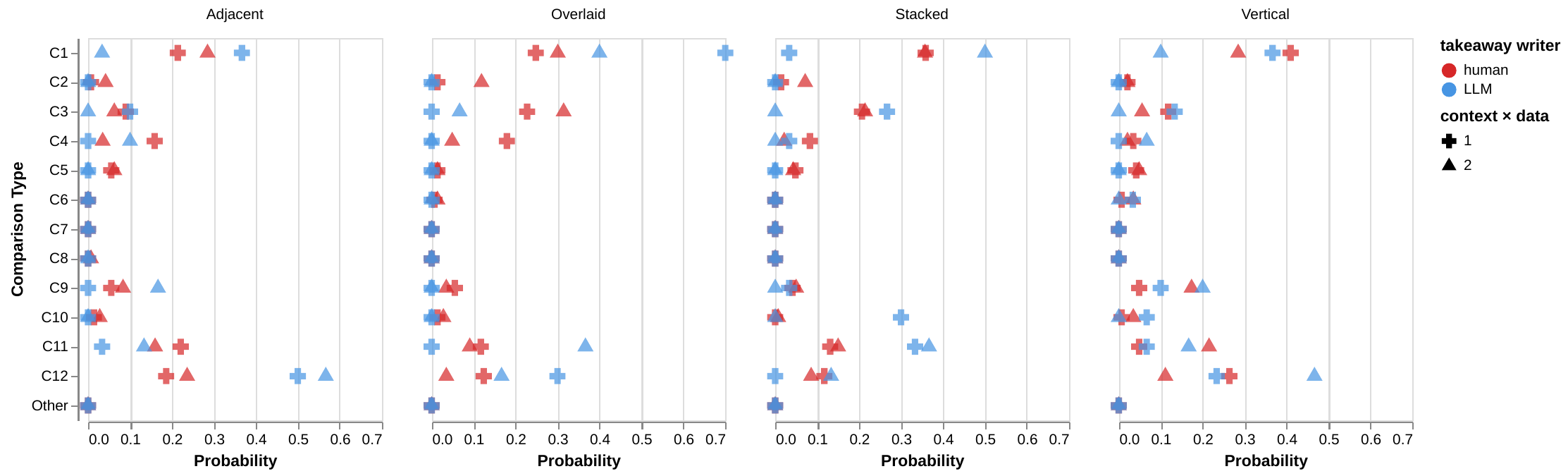}
        \caption{Distributions of human and LLM comparison types for the two example charts for each layout. Unlike humans, whose distributions remain relatively stable across charts using the same layout but different contexts and data, LLM distributions show significant differences.} %\jane{Is this chart showing one-shot and zero-shot LLMs? They don't seem differentiated in the color legend, so unclear what to look for.}}
        
        \label{fig:humanllmexample}
    \end{figure*}
}

\newcommand{\humanzeroone}{
    \begin{figure*}[ht]
        \centering
        \includegraphics[width=\textwidth]{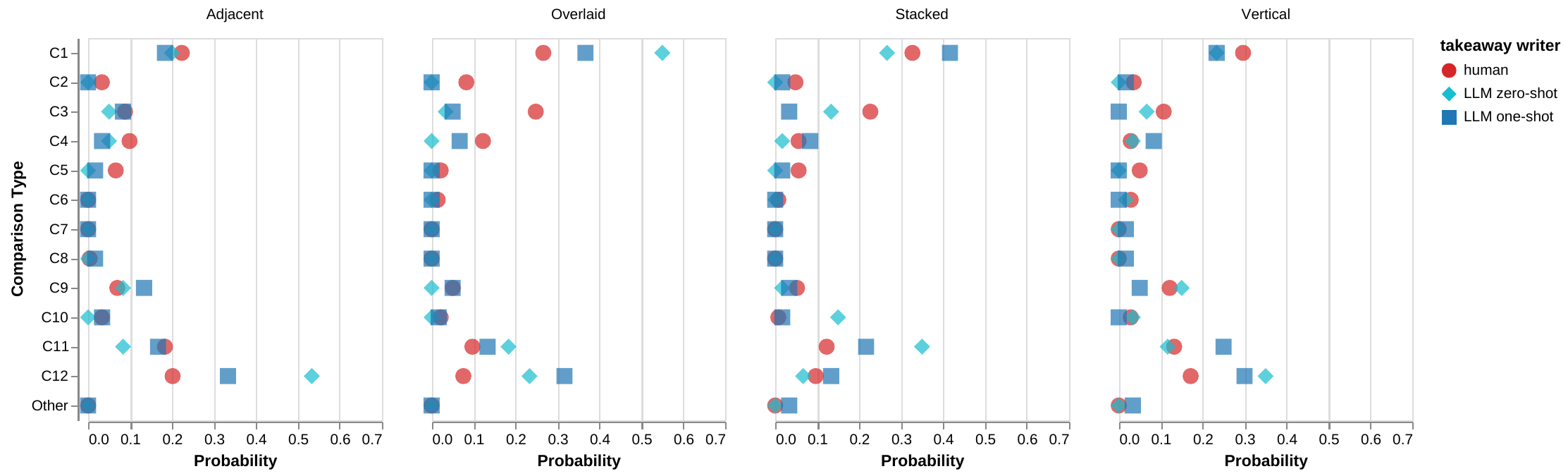}
        \caption{Distributions of human, LLM zero-shot, and LLM one-shot comparison types for each layout. \add{LLM zero-shot distributions are generally closer than one-shot distributions to human ones.}}
        
        \label{fig:humanzeroone}
    \end{figure*}
}

\newcommand{\comparisonCount}{
    \begin{figure}[t]
    \centering
        \includegraphics[width=\columnwidth]{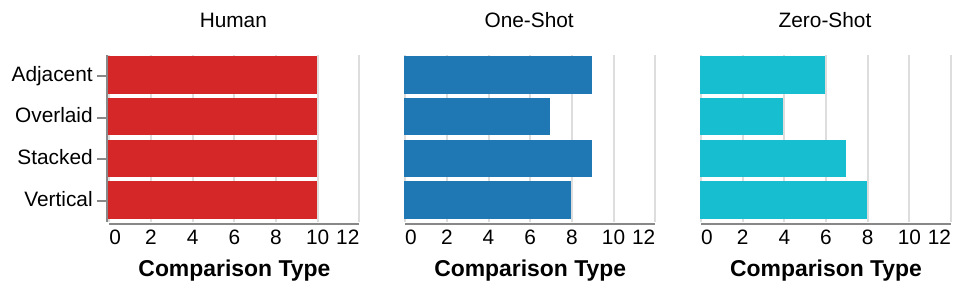}
        \caption{\add{The number of unique comparison types (excluding the ``Other'' category) generated by humans, LLM one-shot, and LLM zero-shot settings. The one-shot setting generates more unique comparison types than the zero-shot setting.}}
        \label{fig:lineup}
        \vspace{-3mm}
    \end{figure}
}

\newcommand{\fourarrangements}{
    \begin{figure}[t]
     \includegraphics[width=\columnwidth]{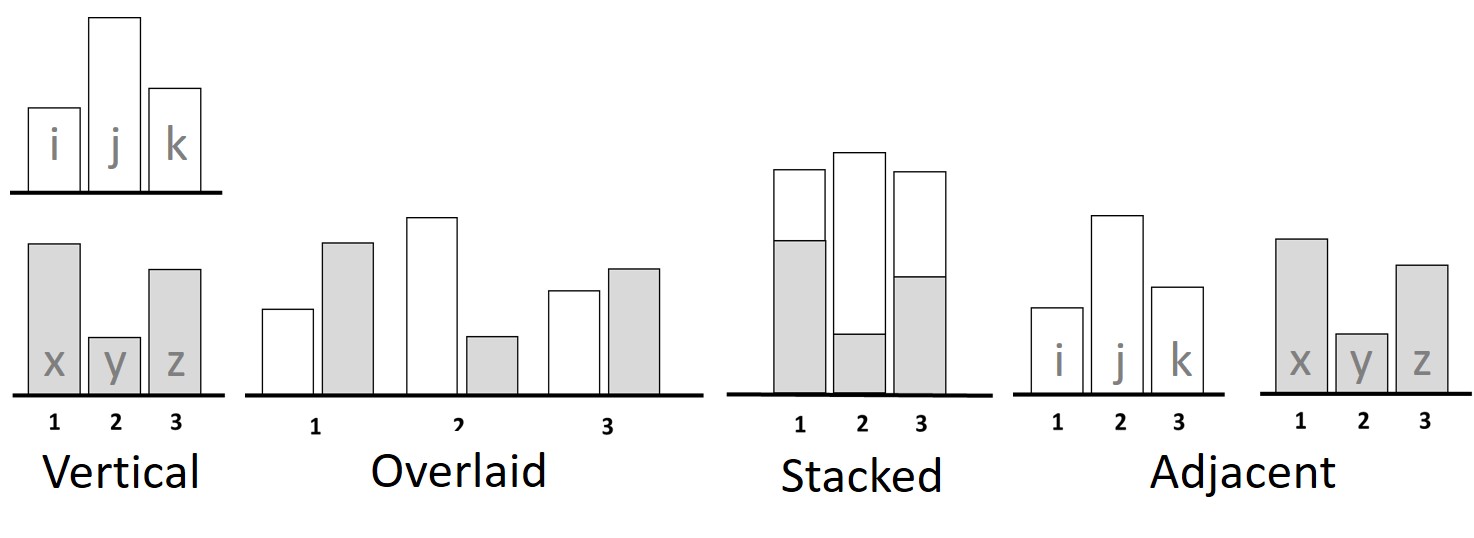}
     \vspace{-18px}
     \caption{Figure from Xiong~\etal~\cite{xiong2021visual} showing four spatial arrangements.}
     \label{fig:4designs}
    \end{figure}
}

\begin{document}

%%%%%%%%%%%%%%%%%%%%%%%%%%%%%%%%%%%%%%%%%%%%%%%%%%%%%%%%%%%%%%%%
%%%%%%%%%%%%%%%%%%%%%% START OF THE PAPER %%%%%%%%%%%%%%%%%%%%%%
%%%%%%%%%%%%%%%%%%%%%%%%%%%%%%%%%%%%%%%%%%%%%%%%%%%%%%%%%%%%%%%%

\firstsection{Introduction}

\maketitle

%% \section{Introduction} %for journal use above \firstsection{..} instead
Designing a visualization involves a series of decisions, ranging from choosing the chart type and selecting a color palette, to determining the amount of accompanying text. 
The data visualization literature has shown that these design choices can significantly influence how readers perceive and interpret the data presented. 
For instance, when presenting the same data, visualizations with higher levels of data aggregation, like bar charts, often lead readers to perceive stronger causal relationships than visualizations with lower levels of aggregation, such as line charts and scatter plots~\cite{xiong2019illusion}. 
While bar charts more readily afford discrete comparisons between data points, line charts are better suited for identifying trends~\cite{zacks1999bars}. Moreover, choosing a suitable visualization format can even help mitigate confirmation bias~\cite{xiong2022reasoning}.

Even when the same chart type is used to visualize a dataset, viewer perception and takeaways may diverge based on other design choices. For bar charts, factors like ordering, partitioning, and spacing are all tied to affordances of different message types~\cite{fygenson2023arrangement}. Different spatial layouts of bars also afford different types of comparisons~\cite{gaba2022comparison, bearfield2023does}. 
For instance, when bars are overlaid (Figure~\ref{fig:4designs}), readers are more likely to identify the maximum and minimum values. However, when bars are adjacently aligned, readers become more likely to pick one bar and compare it to multiple other values.
The richness of design choices and their intricate relations to perceptual and cognitive affordances make designing effective visualizations challenging. 
Even visualization experts can fail to match visualization designs to affordance types in a highly constrained design space of four bar layout possibilities~\cite{xiong2021visual}. 

Large language models (LLMs) have recently taken the world by storm due to their remarkable ability to generate coherent text and follow instructions. 
Trained on vast text corpora, LLMs have the potential to encode rich visualization knowledge. Thus, they~offer exciting new possibilities for visualization tasks. 
For instance, Chen~et~al.~\cite{chen2022nl2interface} utilized Codex~\cite{chen2021evaluating} to transform natural language queries into structurally parameterized SQL, which is used to generate interactive visualization surfaces. 
More powerful LLMs further simplify visualization generation. 
For example, GPT-4 allows users to specify chart requirements and directly produces high-quality visualizations in many instances~\cite{bubeck2023sparks}. 

So can LLM effectively leverage its knowledge base to account for visualization design affordances?
In this paper, we evaluate the sensitivity of LLMs to changes in visualization design compared to humans, when generating visualization takeaways.
%targeted to particular design decisions between charts of the same type. 
\add{Specifically, we explore whether LLMs are perceptually aware of design manipulations like humans are by examining bar charts as a case study.
Existing work by Xiong~\etal~\cite{xiong2021visual} has examined in detail how common layouts affect the type of comparisons people make when reading bar charts. 
The researchers generated two-by-three bar charts using two datasets and visualized them in four layouts (vertically juxtaposed, horizontally juxtaposed, overlaid, and stacked, as shown in Figure~\ref{fig:4designs}).
They collected a corpus of natural language takeaways via crowdsourcing, classifying the types of comparisons human readers made upon seeing these bar charts into 12 categories.
Depending on the spatial layout of the bar chart, a human reader can be more or less likely to make a comparison of a specific type. 
For example, for a dataset depicting the revenue of three stores (1, 2, and 3) in two companies (A and B), people were more likely to compare the revenue of Company A's Store 1 to that of Company B's Store 1 when reading an adjacent bar chart, as shown in Figure~\ref{fig:revenue}, but they do not make that comparison when they read an overlaid or stacked version of the bar chart. 
}

\fourarrangements

\add{We leveraged the experimental results from Xiong~\etal~\cite{xiong2021visual} by asking state-of-the-art LLMs to generate natural language takeaways to the same set of bar charts.
Through three experiments, we compared the distributions of human takeaways from~\cite{xiong2021visual} across the 12 categories to the distributions of LLM takeaways to investigate whether LLMs are perceptually sensitive to changes in visualization designs.
}

\summary

\vspace{6px}\noindent\textbf{Experiment 1: Identifying the optimal configurations.} We experimented with different configurations for generating chart takeaways across four LLMs, two temperatures, nine chart specifications, and two prompting strategies in zero-shot and one-shot settings. We identified representing charts with ggplot2, utilizing GPT-4 with a temperature of 0, and employing the guided discovery strategy as the optimal configuration for the zero-shot setting and representing charts with Vega-Lite and image, using GPT-4V with a temperature of 0.7, and employing the baseline strategy as the optimal configuration for the one-shot setting. In addition, we found that state-of-the-art LLMs can struggle to generate semantically diverse and factually accurate takeaways. 

\vspace{1mm}\noindent\textbf{Experiment 2: Hhuman-LLM-alignment of generated takeaways.} Using the optimal configurations identified in Experiment~1, we generated 30 chart takeaways across four bar chart layouts using two datasets. We found that LLMs, unlike humans, are generally insensitive to bar layouts, while providing an in-context example helps align model generations in some cases.

\vspace{1mm}\noindent\textbf{Experiment 3: Effect of context and data.} \add{If LLMs are not perceptually sensitive to bar layouts, what are they sensitive to?} We observed that the comparison types in LLM takeaways vary greatly between charts with the same layout but different contexts and data values. We showed that humans do not exhibit this sensitivity, thus~revealing inconsistency as another weakness of LLMs. We further demonstrated that context affects comparison types more than data does for most layouts.

\vspace{1mm} Our work is a first step towards understanding the perceptual awareness of LLMs for visualization tasks. In our case study \add{on bar charts}, we not only discovered that LLMs are generally incapable of writing human-aligned takeaways, but they sometimes also generate semantically repetitive and factually incorrect takeaways. \add{We hope our work can motivate investigation into other visualization types and additional dimensions of perceptual awareness.} By exposing the weaknesses of LLMs in reading charts, our work outlines challenges in employing LLMs for visualization tasks and informs future LLM development.

\section{Related Work}

Design choices can influence what patterns people see in data\cite{tversky2014visualizing, few2006information, saket2018task, mackinlay1986automating, nothelfer2019measures}.
For example, showing data as a bar chart is more likely to elicit discrete comparisons (e.g., A is larger than B), whereas a line graph facilitates the detection of temporal trends (e.g., X fluctuates as time passes)~\cite{zacks1999bars, shah2011bar, xiong2019illusion, bearfield2023does}.
% Further, the patterns people see inform how they understand and make decisions with data. 
% For example, chart types that aggregate data points, such as bar charts, increase the likelihood that viewers infer causality from data~\cite{xiong2019illusion}. Conversely, visualizations that show probabilistic outcomes as discrete objects promote a better understanding of uncertainties~\cite{kay2016ish, hawley2008impact, Tait2010, garcia2009communicating}. 
Understanding visual affordances can inform visualization tool design~\cite{gaba2022comparison}. %, such as natural language interfaces~\cite{gaba2022comparison}.
For example, a visualization tool can incorporate affordance mappings as rules to recommend designs that help an analyst see the ``right'' pattern in the data~\cite{zeng2023too}. 
These mappings also support the automatic generation of appropriate captions for visualizations~\cite{bromley2023difference, kim2021towards}. However, the process of identifying visualization affordances across chart types, datasets, design features, and user expertise can be extremely time- and resource-demanding~\cite{fygenson2023arrangement, xiong2021visual, boy2015suggested}. It is also challenging to incorporate such rules into visualization systems given the diverse possible use cases and syntactic and semantic variations in users' natural language input.

\vspace{-6px}
\add{\subsection{LLMs for Visualization}}
\noindent \add{LLMs offer exciting possibilities for the visualization community~\cite{schetinger2023doom, ye2024generative}. Not only do they provide a unified natural language interface for a suite of visualization tasks, but they also have the potential to encode visualization knowledge and best practices due to their exposure to Internet-scale training data. A natural application of LLMs for visualization is natural language to visualization (NL2VIS). For instance, Wu~\etal~\cite{wu2024automated} explored using LLMs for NL2VIS and achieved state-of-the-art performance. Tian~\etal~\cite{tian2024chartgpt} integrated LLMs into a step-by-step reasoning pipeline capable of generating visualizations from abstract natural language.}

\add{There is also an increasing body of work that explores using LLMs to generate natural language utterances for visualizations.} For example, Tang et al.~\cite{tang2023vistext} assembled a dataset of human-written visualization captions and provided initial results on generating captions with a vision-language model. Ko~\etal~\cite{ko2023natural} generated natural language datasets from Vega-Lite charts using GPT-4 with some success. \add{In addition, LLMs have been used to generate data presentations~\cite{wang2024outlinespark}, enhance visual analytics~\cite{zhao2024leva}, and guide chart reading~\cite{choe2024enhancing}.} However, to the best of our knowledge, no work so far has assessed the extent to which LLMs are perceptually aware in visualization tasks like humans are. This motivates us to contribute a case study as an initial step toward understanding how well LLMs can predict human takeaways from visualizations.

\add{\subsection{Comparison Types in Bar Charts}}
\label{relatedworkXiong}
The spatial layout of data can change which data points people compare and what they take away~\cite{xiong2021visual, fygenson2023arrangement}. 
For example, as shown in Figure~\ref{fig:4designs}, when people look at bar charts in the adjacent layout, they tend to pick out one bar and compare it to multiple other bars. 
If the bars are in a vertical layout, people would instead compare the vertically aligned bars. 
Xiong~\etal~\cite{xiong2021visual} described 12 type of comparisons people engage in when reading bar charts. 
These categories are grounded and refined based on empirical data~\cite{gaba2022comparison}.
We describe the 12 categories of comparisons below to contextualize the current work. 
We first introduce the concept of \textbf{cardinality}, which has to do with how people group data values prior to making comparisons. 
The 12 comparison categories can be divided into four cardinalities:
\vspace{1mm}

\noindent \textbf{one-to-one (C1,C5, C9)}: Comparison between one bar and another bar.
\vspace{1mm}

\noindent \textbf{two-to-two (C2, C6, C10))}: Comparison between one set of two bars and another set of two bars.
\vspace{1mm}

\noindent \textbf{all elements (C3, C7, C11)}: Comparison between one set of bars and the remaining set of bars.
\vspace{1mm}

\noindent \textbf{one-to-multiple (C4, C8, C12)}: Comparison between one bar and a set of bars.
\vspace{1mm}

Each cardinality consists of three types of comparison approaches.
Referencing the example from Figure~\ref{fig:revenue}, we refer to the two companies, A and B, as two ``groups'', and each of the three data points associated with the store revenues within each group as ``elements''.
\vspace{1mm}

\noindent \textbf{Across group - Within element (C1, C2, C3, C4):} The reader identifies the same element(s) in each group and compares them. 
\vspace{1mm}

\noindent \textbf{Across group - Across element (C5, C6, C7, C8):} One element from one group is compared to a different element in another group. 
\vspace{1mm}

\noindent \textbf{Within group - Across element (C9, C10, C11, C12):} The reader identifies one group and compares different elements within that group.

In general, amongst the popular comparison types, C1 and C11 are commonly compared for all bar layouts. C3 and C4 are especially associated with the overlaid layout. C9 is commonly made when readers see a vertical bar chart.
More detailed distributions of comparison types can be found in Xiong~\etal~\cite{xiong2021visual}.

\section{Experiment 1: Optimal Configurations}
\label{sec:experiment1}
%%%%%%%
%% Note: Comment in one of the following labels for the section to keep track of progress
%%%%%%%
% \complete
% \feedbackProvided
% \readyForFeedback
% \underRevision
% \incomplete
%%%%%%%
%% Note: Add the content for the section directly after this comment
%%%%%%%

The appropriateness of chart takeaways generated by LLMs is informed by several parameters, including (1)~the choice of which LLM to use~(\eg GPT-3.5, GPT-4, and Llama 2~\cite{touvron2023llama2}), (2)~the LLM temperature setting, which dictates the randomness in the model's responses,
(3)~the input chart specification (\ie how the chart is represented), and (4)~the prompting strategy (\ie how natural language prompts are used to instruct the LLM to generate chart takeaways). 

In addition to these four parameters, research has shown that providing LLMs with in-context examples can boost their performance~\cite{brown2020language}. To this end, we tested LLM performance in two settings: zero-shot, where we requested takeaways without any example chart or takeaways, and one-shot, where we provided the LLMs with a sample chart and human takeaways before requesting takeaways on the test case.

In order to generate human-aligned chart takeaways, we first need to ensure that the generations are accurate. In addition, since humans tend to write diverse takeaways for any given chart, an optimal configuration must be capable of generating semantically diverse takeaways. In Experiment 1, we sought to determine a configuration for the four parameters to optimize both takeaway accuracy and semantic diversity. Using stimuli from Xiong~\etal~\cite{xiong2021visual}, we partially replicated their study by prompting LLMs to produce 30 semantically diverse takeaways and assessed takeaway accuracy and semantic diversity for visualizations.

\subsection{Model Types and Temperature Settings}
\label{sec:model}
%%%%%%%
% \complete
% \feedbackProvided
% \readyForFeedback
% \underRevision
% \incomplete
%%%%%%%
We generated takeaways using four state-of-the-art LLMs at the time of experiment: \add{GPT-4-1106-vision-preview}\footnote{https://platform.openai.com/docs/models/gpt-4-and-gpt-4-turbo} (hereafter GPT-4V), \add{GPT-4-0613}\footnote{https://platform.openai.com/docs/models/gpt-4-and-gpt-4-turbo} (hereafter GPT-4), \add{GPT-3.5-turbo-1106}\footnote{https://platform.openai.com/docs/models/gpt-3-5} (hereafter GPT-3.5), and Gemini~1.0~Pro\footnote{https://deepmind.google/technologies/gemini/\#gemini-1.0}. 
For each model type, we experimented with two temperature settings. The temperature values were set at one of \{0, 0.7\}. A lower temperature value results in more focused and deterministic output, while a higher value allows for more randomness and creativity in the responses. We chose these two values because a temperature of 0 corresponds to greedy decoding and minimizes randomness, whereas a value of 0.7 is a typical default setting for many LLMs.

\subsection{Datasets and Chart Specifications}
%%%%%%%
% \complete
\feedbackProvided
% \readyForFeedback
% \underRevision
% \incomplete
%%%%%%%
Xiong~\etal~\cite{xiong2021visual} generated two datasets for their stimuli, each containing two groups of three data points. 
%\will{do we need to further describe the datasets? they don't really have names and are extremely simple--just 6 data points.} 
They visualized the data using bar charts and tested four spatial layouts---vertically juxtaposed, horizontally juxtaposed, overlaid, and stacked (see Figure~\ref{fig:4designs})---resulting in eight visualizations. Figure~\ref{fig:revenue} shows an example visualization using the adjacent layout to depict the revenues of two companies (Company~A and Company B) across three stores (Store 1, Store 2, and Store 3).

To generate chart takeaways using LLMs, we must represent charts in a way that can be recognized by the models. Inspired by Tang~\etal~\cite{tang2023vistext}, we experimented with three classes of chart representations in this work: text-based, image-based, and a combination of both text and image. For text-based representations, we explored four chart representations: Vega-Lite~\cite{satyanarayan2016vega}, Matplotlib~\cite{hunter2007matplotlib}, ggplot2~\cite{wickham2016ggplot2}, and scene graphs, which are hierarchical representations of the visual elements in a visualization. In order to authentically reproduce Xiong~\etal~\cite{xiong2021visual}, we made sure that these textual representations, when rendered as images, looked similar to the original stimuli that human participants saw.
For image-based representations, we simply represented visualizations as bitmap images. In addition, we tried combining one of the text-based representations with images. For instance, in one experiment, we fed both the image (Figure~\ref{fig:revenue}) and the Matplotlib specification producing it to GPT-4V. In total, we considered nine chart representations (four text-based, one image-based, and four combined). We include all eight stimuli and their specifications in the supplemental materials.

\subsection{Prompting}
\label{sec:prompting}
%%%%%%%
% \complete
\feedbackProvided
% \readyForFeedback
% \underRevision
% \incomplete
%%%%%%%
Prompting is essential to the performance of LLMs (e.g.,~\cite{yao2023tree},~\cite{wei2022chain}). Good prompts employ appropriate strategies and articulate the tasks clearly. In this work, we explored two prompting strategies: the baseline strategy and the guided discovery strategy~\cite{ko2023natural} (detailed in Section~\ref{sec:guided}). To understand what language to use for system prompts \add{(instructions guiding the behavior of the LLM)} and user prompts \add{(specific user queries)}, we piloted several prompts on GPT-4 using the baseline strategy but with slight variations in wording. For instance, we tested system prompts with different levels of specificity, such as "You are a helpful research~assistant" and "I am a visualization researcher and you are a helpful research assistant. You should draw on your knowledge of graphical perception research to predict what humans will write as takeaways to visualizations." We adopted the task description and system prompt yielding takeaways with good semantic diversity and high accuracy, and adapted them to the prompting strategies. 

\subsubsection{Baseline Strategy} This approach begins with a straightforward framing of the task, followed by the chart representation. See a zero-shot example below: 

\vspace{1em}

\begin{mdframed}
\footnotesize
\noindent\texttt{System Prompt: I am a visualization researcher and you are a helpful research assistant. You should draw on your knowledge of graphical perception research to predict what humans will write as takeaways to visualizations.\\[1em]
User Prompt: I will show you some code below, which generates a chart. The chart depicts the revenue of three stores selling computers from two companies. The three stores are Store 1, Store 2, and Store 3, and the two companies are Company A and Company B. Your job is to predict what humans will write as takeaways. Note that humans only see the visualization. Since you are a language model, I will show you the code producing the visualization. Remember, your takeaway should not only be a description of the chart; it should encapsulate a take-home message from the chart. Generate 30 semantically different takeaways for this chart. It's okay for takeaways to not be full sentences.\\[1em]
\gray{<< Chart specification omitted here for conciseness. >>}}
\end{mdframed} 

% \vspace{1em}
\figureRevenue

\subsubsection{Guided Discovery} 
\label{sec:guided}
The guided discovery strategy was proposed by Ko~\etal~\cite{ko2023natural} in their framework generating natural language datasets from Vega-Lite specifications and drew from chain-of-thought prompting~\cite{wei2022chain} and educational psychology~\cite{brown1994guided}. 
In guided discovery, the user provides scaffolding and poses key questions in the prompt to help the model reason and extract insights. 
We instructed the model to reason step-by-step: first, we asked the model to extract the chart type and the variables depicted for grounding purposes (scaffolding); then, we instructed the model to attend to bar heights and reason about the pattern shown; finally, we requested 30 semantically distinct takeaways from the visualization and included the chart representation.
See \add{the following zero-shot} example: 

% \vspace{1em}

\begin{mdframed}
\footnotesize
\noindent\texttt{\noindent\gray{<< System prompt omitted here for conciseness. >>}\\[1em]
User Prompt:  I will show you some code below, which generates a chart. The chart depicts the revenue of three stores selling computers from two companies. The three stores are Store 1, Store 2, and Store 3, and the two companies are Company A and Company B. Your job is to predict what humans will write as takeaways. Note that humans only see the visualization. Since you are a language model, I will show you the code producing the visualization.\\[1em]
\gray{<< Chart specification omitted here for conciseness. >>}\\[1em]
Let's think step by step. What type of chart is this? What are the variables depicted?\\[1em]
Assistant: \gray{<< Response omitted. >>}\\[1em]
User Prompt: Look at the relative heights of the bars. Briefly, what patterns do the bars show? \\[1em]
Assistant: \gray{<< Response omitted. >>}\\[1em]
User Prompt: Now, generate 30 semantically different takeaways for this chart. Remember, your takeaway should not only be a description of the chart; it should encapsulate a take-home message from the chart. It's okay for takeaways to not be full sentences. 
}
\end{mdframed} 

% \vspace{1em}

\subsubsection{Prompts for one-shot settings}

 The examples in the previous sections are instantiations of templates for zero-shot settings. For one-shot settings, we provided a sample bar chart using the same spatial layout and specified in the same format as the test case and approximately 150 human takeaways before asking the model for takeaways on the test case. We also tested both the baseline and the guided discovery strategies for one-shot prompts. We largely adhered to the task descriptions and system prompts previously identified. Please refer to the supplemental materials for our one-shot~templates.

\subsection{Experiment 1: Procedure and Setup}
\label{sec:procedure1}
%%%%%%%
% \complete
\feedbackProvided
% \readyForFeedback
% \underRevision
% \incomplete
%%%%%%%
The main goal of Experiment 1 is to identify optimal configurations of the four parameters outlined in Section~\ref{sec:experiment1} for zero-shot and one-shot settings. To this end, we prompted LLMs to generate 30 semantically diverse takeaways for each input chart specification. As discussed in the previous sections, we evaluated four LLMs, two temperatures, nine chart representations\footnote{Note that we tested chart specifications involving bitmap images only on GPT-4V since it was the only LLM with vision capability among the four tested.}, and two prompting strategies on two datasets in both zero-shot and one-shot settings. Given the large number of configurations, we only experimented on the adjacent layout in Experiment 1 to derive optimal configurations. This resulted in a total of 272 trials across both zero-shot and one-shot settings. To ensure the integrity \add{and independence} of our results, we conducted each trial \add{through separate API calls}. Finally, we evaluated the takeaways according to the procedure in Section~\ref{eval1} and identified the optimal configurations.

\subsection{Experiment 1: Evaluation Approaches}
\label{eval1}
%%%%%%%
% \complete
\feedbackProvided
% \readyForFeedback
% \underRevision
% \incomplete
%%%%%%%
We evaluated the takeaways from each configuration on two metrics: semantic diversity and factual accuracy. 

\clusters

\subsubsection{Semantic Diversity} 
\label{semantic diversity}

Semantic diversity measures the semantic variety of takeaways, \add{via a ``cluster count'' proxy. We passed the takeaways for each configuration to GPT-4 and instructed it to cluster them based on their semantics.} Take Figure~\ref{fig:revenue} as an example, the takeaway ``the revenue of Company A at Store 3 is less than that of Company B'' is semantically equivalent to ``the revenue of Company B at Store 3 is more than that of Company A''. \add{Therefore, we consider these takeaways to be in the same cluster. These takeaways mean different things from ``Store 2 generates the highest revenue for Company A'', and thus we consider them to be from different clusters.}
\add{Hence, for a set of takeaways, the higher the cluster count, the more semantically diverse it is.}

Since the prompts request 30 semantically distinct takeaways, we first filtered out configurations that generated semantically repetitive takeaways. Recognizing that in-context examples could influence the generated outcomes, we conducted the filtering process separately for zero-shot and one-shot settings. 
% To evaluate semantic diversity, we first passed the takeaways from each configuration to GPT-4 and instructed it to cluster them based on their meaning. We proxied semantic diversity using the cluster count returned by GPT-4---the higher the cluster count, the more semantically diverse a set of takeaways.
To verify that the cluster counts returned by GPT-4 are reasonable measures of semantic diversity, we selected 10 configurations at the $10^{th}$, $20^{th}$, \emph{...}, and $100^{th}$ percentile of cluster count, manually clustered them, and compared machine and human-generated cluster counts.

Next, we created two rankings for all configurations based on semantic diversity as determined by GPT-4, one each for the zero-shot and one-shot settings. We only reviewed configurations generating takeaways in the top 25\% in terms of semantic diversity for accuracy of their takeaways and discarded the rest for insufficient semantic diversity.

\subsubsection{Factual Accuracy}

Factual accuracy measures if takeaways accurately reflect information in the visualization. Two coders manually coded all takeaways for factual accuracy from the top 25\% of configurations in terms of semantic diversity. Initially, the two coders separately coded 30 model-generated takeaways for a given visualization. Next, they discussed and established guidelines to categorize the takeaways until reaching a consensus on which ones were accurate, ambiguous, and inaccurate. For example, in Figure~\ref{fig:revenue}, ``Store 2 generates much higher revenue than Store 1 does for Company A'' is an accurate takeaway, ``the strategy a company uses affects its revenue'' is an ambiguous takeaway, and ``Store 1 generates less revenue than does Store 3 for Company B'' is an incorrect takeaway. 

The two coders each coded half of the takeaways and calculated the percentages of correct, ambiguous, and inaccurate takeaways for each configuration. 
Since we used each configuration to generate takeaways for two charts, one coder coded each chart and the results from both coders were averaged to obtain the factual accuracy score for that configuration. This approach counterbalances inter-rater differences. 

A set of takeaways is considered more factually accurate if it has a higher percentage of correct takeaways compared to another set. When the percentage of correct takeaways is equal, the set with the lower percentage of factually inaccurate takeaways is deemed more accurate.

\subsection{Experiment 1: Results}
%%%%%%%
% \complete
% \feedbackProvided
% \readyForFeedback
% \underRevision
% \incomplete
%%%%%%%

In this section, we first confirm that GPT-4's clustering results are reasonable proxies for semantic diversity. Then, we present results on the semantic diversity of takeaways produced by each configuration. Finally, we report the optimal configurations for zero-shot and one-shot settings based on factual accuracy.

\subsubsection{Confirming GPT-4's Clustering Results}

We first verified that the cluster counts generated by GPT-4 are reasonably accurate. After sampling ten sets of takeaways with a diverse range of GPT-4-generated cluster counts following the procedure outlined in Section~\ref{semantic diversity} and manually clustering them, we performed linear regression to assess the relationship between the GPT-4-generated cluster counts and the manually obtained cluster counts. Results indicated that GPT-4 cluster counts indeed predict manual cluster counts quite well ($r^2 = 0.83$, $p < 0.001$). Hence, we concluded that the cluster counts generated by GPT-4 are good proxies for semantic diversity.

\subsubsection{Semantic Diversity by Configuration}

All configurations tested successfully generated 30 takeaways for each visualization. We visualize the distribution of cluster counts as determined by GPT-4 for every set of zero-shot and one-shot takeaways in Figure~\ref{fig:cluster counts}, broken down by the LLM type. In either case, \textbf{there is a wide range of semantic diversity across configurations}.

After retaining the top 25\% of configurations as described in Section~\ref{semantic diversity}, we are left with configurations with an average cluster count on the two sets of takeaways of above 21 in the zero-shot setting and those with an average above 19.5 in the one-shot setting. This procedure left us with 19 configurations to code for accuracy in the zero-shot setting and 18 in the one-shot setting. In both settings, Gemini~1.0~Pro produces the highest number of semantically diverse takeaways.

% \zeroshotclusters

% \oneshotclusters

\subsubsection{Accuracy by Configuration}

\begin{table}[t]
    \caption{Top five configurations in terms of factual accuracy in the zero-shot setting and their average number of accurate~(\cmark), ambiguous~(\textbf{\textit{?}}\,), and inaccurate~(\xmark) takeaways across the two visualizations tested.}
    \centering
    \begin{tabu} to \textwidth {ccccccc}
       \toprule
       LLM & Temp. & Chart Spec. & Prompt & \cmark & \textbf{\textit{?}} & \xmark \\
       \midrule
       GPT-4 & 0 & ggplot2 & CoT & 23 & 5 & 2 \\ \midrule
       GPT-4V & 0.7 & VL+image & baseline & 21.5 & 6 & 2.5 \\ \midrule
       GPT-4V & 0.7 & matplotlib & baseline & 21 & 6.5 & 2.5 \\ \midrule
       GPT-4V & 0 & scene+image & CoT & 20.5 & 8 & 1.5 \\ \midrule
       GPT-4 & 0 & ggplot2 & baseline & 20 & 2 & 8 \\
       \bottomrule
    \end{tabu}
    \label{tab:zero}
\end{table}

\begin{table}[t]
    \caption{Top five configurations in terms of factual accuracy in the one-shot setting and their average number of accurate~(\cmark), ambiguous~(\textbf{\textit{?}}\,), and inaccurate~(\xmark) takeaways across the two visualizations tested.}
    \centering
    \begin{tabu} to \textwidth {ccccccc}
       \toprule
       LLM & Temp. & Chart Spec. & Prompt & \cmark & \textbf{\textit{?}} & \xmark \\
       \midrule
       GPT-4V & 0.7 & VL+image & baseline & 24 & 4.5 & 1.5 \\ \midrule
       GPT-4 & 0 & Vega-Lite & CoT & 24 & 1.5 & 4.5 \\ \midrule
       GPT-4V & 0 & scene+image & CoT & 23.5 & 2 & 4.5 \\ \midrule
       GPT-4 & 0 & matplotlib & baseline & 22.5 & 2 & 5.5 \\ \midrule
       GPT-4 & 0 & scene graph & baseline & 22 & 2 & 6 \\
       \bottomrule
    \end{tabu}
    
    \label{tab:one}
\end{table}

Table~\ref{tab:zero} and Table~\ref{tab:one} show the configurations in the top five in terms of average factual accuracy for the zero-shot and one-shot settings, respectively. While there are a mix of temperatures, chart representation, and prompting strategies in the top five, all configurations are notably either generated by GPT-4 or GPT-4V, which suggests \textbf{the GPT-4 family's dominance over other LLMs in generating accurate takeaways}. We also find that Gemini~1.0~Pro, while capable of writing semantically diverse takeaways, struggles in factual accuracy. In fact, most configurations involving Gemini~1.0~Pro average below 33.3\% in accuracy and the majority of takeaways are ambiguous. Also~underperforming in accuracy is GPT-3.5, which often produces more factually inaccurate takeaways than accurate ones. Overall, our results show that \textbf{generating accurate chart takeaways is still a challenging task for~LLMs}.

Following the procedure in Section~\ref{eval1}, we identified representing charts with ggplot2, utilizing GPT-4 with a temperature of 0, and employing the guided discovery strategy as the optimal configuration for the zero-shot setting and representing charts with Vega-Lite and image, using GPT-4V with a temperature of 0.7, and employing the baseline strategy as the optimal configuration for the one-shot setting.

\add{Due to the black-box nature of LLMs, it is challenging to determine why certain configurations outperformed others. Nonetheless, our results suggest that the choice of LLM has the largest impact on the quality of the generated takeaways. While different temperature values, chart specifications, and prompting strategies appear in the top-performing configurations, the model types remain surprisingly consistent, underscoring the importance of choosing the right LLM for generating takeaways. This type of qualitative performance improvement resulting from using a superior LLM has been similarly observed across many tasks~\cite{bubeck2023sparks}.}

%!TEX root = proceedings.tex

\section{Experiment 2: Alignment of Human \& LLM Takeaways}
%%
%% Note: Comment in one of the following labels for the section to keep track of progress
%%
    %\complete
    %\feedbackProvided
    %\readyForFeedback
    %\underRevision
    % \incomplete
%%
%% Note: Add the content for the section directly after this comment
%%
In Experiment 1, we generated takeaways from adjacent bar charts and identified optimal configurations for the zero-shot and one-shot settings. Next, we started to more generally understand whether LLMs are influenced by the spatial layouts of bars like humans are. We tested all four layouts in Figure~\ref{fig:4designs}, using the optimal configurations identified in Experiment 1. We prompted LLMs to generate 30 takeaways for each visualization as if it were 30 people in both zero-shot and one-shot settings. In addition to coding takeaways for accuracy, we classified them along the taxonomy Xiong~\etal developed, which captures 12 types of comparison each takeaway makes (denoted C1, C2, \emph{...}, C12, see Section 2.2). 
%Since all stimuli in Xiong~\etal depict two groups, each with three data points, people naturally tend to make comparisons in their takeaways. 
%One dimension to categorize takeaways is whether they compare data \textit{within a group} or \textit{across groups}. Another dimension is \textit{the numbers of data points involved in the comparison}. 
% In total, this taxonomy includes twelve comparison types (denoted C1, C2, \emph{...}, C12). 
By comparing the distributions of LLM-generated takeaways vs. human-written takeaways along these comparison types, we can understand to what extent takeaways from LLM emulate those from humans. 
% whether takeaways from LLMs are influenced by the spatial layouts of bars like humans are. 

\subsection{Experiment 2: Procedure and Setup}

We used the same two datasets as in Experiment 1. We started by replicating all eight original stimuli across the four bar chart layouts from Xiong~et~al. in ggplot2 and Vega-Lite, which were optimal chart specifications for zero-shot and one-shot settings identified in Experiment~1. We then generated takeaways for every stimulus. In the zero-shot setting, we represented charts using ggplot2, adopted GPT-4 as the LLM, set its temperature to 0, and prompted the model following the guided discovery strategy. In the one-shot setting, we represented charts using both Vega-Lite specifications and bitmap images, adopted GPT-4V as the LLM, set its temperature to 0.7, and prompted the model following the baseline strategy. The prompts we used in Experiment 2 were exactly the same as the ones in Experiment 1, except that instead of asking the model to "generate 30 semantically different takeaways for this chart", we asked it to "generate 30 takeaways for this chart as if you were 30 people".

\subsection{Experiment 2: Evaluation Approaches}

We evaluated the takeaways generated by LLMs across two dimensions: factual accuracy and human-LLM alignment of comparison types.

Factual accuracy was measured in the same way as in Experiment 1. One coder coded all charts for factual accuracy. We also leveraged the taxonomy from Xiong~\etal~\cite{xiong2021visual}, and one coder coded all the takeaways for comparison types. Since we observed that some LLM-generated takeaways did not make any comparisons, we added an additional category, ``Other'', to capture them. For each bar chart layout, we identified the top three most common comparison types averaged across the two examples in model-generated takeaways (denoted as the set $M$) and in human-written takeaways (denoted as the set $H$). We calculated the overlap between $M$ and $H$, a metric we will shorthand as precision@3:

\[
\text{precision@3} = \frac{\left| H \cap M \right|}{3}.
\]

To quantify the distances between the distributions of LLM and human comparison types, we \add{first} normalized the frequency distributions into probability distributions ($P_M$ for model and $P_H$ for human) \add{by dividing the number of takeaways falling into each comparison type by the total number of takeaways.} \add{Next, we} calculated Total Variation Distance (TVD) between $P_M$ and $P_H$, which is half of the sum of the absolute differences between the probabilities for each \add{of the thirteen categories (the original twelve categories from Xiong~\etal~\cite{xiong2021visual} and the ``Other'' category)}. The value of TVD ranges between zero and one. The closer TVD is to zero, the closer the two distributions are. More formally, TVD between $P_M$ and $P_H$ is defined~as:

\[
\text{TVD}(P_H, P_M) = \frac{1}{2} \sum_{i=1}^{13} |p_{M_i} - p_{H_i}|,
\]

\subsection{Experiment 2: Results}

In this section, we report how the optimal configurations performed on takeaway accuracy and alignment with human comparison types.

\begin{table}[t]
    \caption{Zero-shot takeaway accuracies for the four layouts. \add{Even state-of-the-art LLMs can struggle to write factually accurate takeaways.}}
    \centering
    \begin{tabu}{%
  	  c%
  	  	*{7}{c}%
  	  	*{2}{c}%
  	}
       \toprule
       Layout & Accurate & Ambiguous & Inaccurate \\
       \midrule
       Adjacent  &  68.33\% & 15.00\% & 16.67\% \\
       Overlaid & 80.00\% & 15.00\% & 5.00\% \\
       Stacked  & 21.67\% & 18.33\% & 60.00\% \\
       Vertical & 56.67\% & 3.33\% & 40.00\% \\
       \bottomrule
    \end{tabu}
    \label{tab:zeroshot}
\end{table}
\begin{table}[t]
    \caption{One-shot takeaway accuracies for the four layouts. \add{One-shot accuracies improve upon zero-shot accuracies in most cases.}}
    \centering
    \begin{tabu}{%
  	  c%
  	  	*{7}{c}%
  	  	*{2}{c}%
  	}
       \toprule
       Layout & Accurate & Ambiguous & Inaccurate \\
       \midrule
       Adjacent  &  70.00\% & 3.33\% & 26.67\% \\
       Overlaid & 65.00\% & 5.00\% & 30.00\% \\
       Stacked  & 51.67\% & 3.33\% & 45.00\% \\
       Vertical & 73.33\% & 0.00\% & 26.67\% \\
       \bottomrule
    \end{tabu}
    \label{tab:oneshot}
    \vspace{-3mm}
\end{table}
\humanzeroone

\subsubsection{Accuracy}

Tables~\ref{tab:zeroshot} and ~\ref{tab:oneshot} show the accuracies of takeaways across zero-shot and one-shot settings, respectively. In the zero-shot setting, accuracies varied significantly across layouts, with a minimum of 21.67\% for the stacked layout and a maximum of 80.00\% for the overlaid layout. We observed that \textbf{in many cases LLMs suffered from significant hallucination}. For instance, when generating zero-shot takeaways for the stacked layout, 60\% of takeaways were factually inaccurate. Upon examining these incorrect takeaways, we found that the LLM consistently confused the group labels. This points to weaknesses even in state-of-the-art LLMs to accurately read and reason about visualizations. 

The one-shot setting yielded more consistent accuracies across different layouts in the range of 51.67\% to 73.33\%. We also found that one-shot accuracies were higher than zero-shot accuracies in three of the four layouts, suggesting \textbf{the benefits of in-context learning on grounding takeaway generation.} In particular, the presence of example takeaways boosted the accuracy for the stacked layout by 30\%. Furthermore, one-shot takeaways tended to contain fewer ambiguous statements. That said, at least 26.67\% of one-shot takeaways in every layout did not accurately portray the visualization. Hence, there is still much room for improvement when it comes to generating accurate chart takeaways with LLMs.

\begin{table}[t!]
    \caption{TVD's between the comparison type distributions of zero-shot takeaways and human takeaways, and between one-shot takeaways and human takeaways for each layout. \add{One-shot TVDs are lower in most cases, suggesting the benefit of providing human-written takeaways in calibrating LLM generations.}}
    \centering
    \begin{tabu}{%
  	  c%
  	  	*{7}{c}%
  	  	*{2}{c}%
  	}
       \toprule
       Layout & Zero-Shot TVD & One-Shot TVD\\ 
       \midrule
        Adjacent & 0.3456 & 0.2090 \\ 
        Overlaid & 0.5278 & 0.3778 \\ 
        Stacked & 0.3699  & 0.2903 \\ 
        Vertical & 0.2167 & 0.3679 \\ 
       \bottomrule
    \end{tabu}
    \label{tab:tvd1}
\end{table}

\subsubsection{Alignment of Comparison Types}

Figure~\ref{fig:humanzeroone} depicts the distributions of comparison types of human-written takeaways, LLM zero-shot takeaways, and LLM one-shot takeaways for the four layouts. Visually, the distributions of zero-shot takeaways tend to deviate from the human takeaways (e.g., zero-shot takeaways place an overly very high probability mass on C12 for the adjacent layout), while the distributions of one-shot takeaways look more similar to human-written takeaways. Table~\ref{tab:tvd1} shows the TVD's between the distributions of comparison types for LLM zero-shot takeaways and human takeaways, and between those for LLM one-shot takeaways and human takeaways. In all but the vertical layout, TVD is lower in the one-shot setting, confirming the visual conclusion. Even though our prompt did not outline alignment in comparison types as an optimization objective, \textbf{the LLM better calibrated the comparison types in its generations when an example was provided}.

Figure~\ref{fig:lineup} shows the number of unique comparison types (excluding the ``Other'' category ) in human, LLM zero-shot, and LLM one-shot takeaways. While human takeaways consistently covered a diverse set of comparison types ($mean~=~10$), the LLMs tended to focus on fewer comparison types across all layouts in the zero-shot setting ($mean~=~6.25$). When provided with example human takeaways, the LLMs showed expanded coverage of comparison types ($mean~=~8.25$), demonstrating the benefits of in-context learning for simulating the diverse takeaways humans tend to write.

\comparisonCount

Table~\ref{tab:top comp types} further details the top three most frequent comparison types for each layout from each source while Table~\ref{tab:precision} shows precision@3. Compared to zero-shot takeaways, one-shot takeaways attained higher precision@3 for two layouts and equal precision@3 for the remaining two, indicating that \textbf{the presence of example takeaways tends to allow LLMs to better capture the most afforded comparison types}. While LLMs were generally good at predicting frequent comparison types across all layouts, such as C1, C11, and C12, when rarer comparison types made it to the top three, such as C3 and C4, LLMs failed to predict them. However, the frequent appearance of a less common comparison type in other layouts within the top ranks for a specific layout suggests a unique affordance by that layout for such comparisons. Failure by the LLM to capture these affordances is further evidence that \textbf{LLM comparison types are generally insensitive to bar layout}.

\newcommand{\cOne}{\small\textcolor{lightorange}{C1}}
\newcommand{\cThree}{\small\textcolor{lightred}{C3}}
\newcommand{\cFour}{\small\textcolor{lightgold}{C4}}
\newcommand{\cNine}{\small\textcolor{lightpink}{C9}}
\newcommand{\cTen}{\small\textcolor{lightaqua}{C10}}
\newcommand{\cEleven}{\small\textcolor{lightpurple}{C11}}
\newcommand{\cTwelve}{\small\textcolor{lightgreen}{C12}}
\begin{table}[t!]
    \caption{The top three most frequent comparison types generated for each layout in the human, LLM zero-shot, and LLM one-shot takeaways. \add{LLMs are good at predicting frequent comparison types across all layouts, but fail to predict when generally rarer types are in the top three.}}
    \centering
    \setlength{\tabcolsep}{3.5pt}
    \begin{tabu}{%
  	  c%
  	  	*{12}{c}%
  	  	*{2}{c}%
  	}
       \toprule
       Layout & & \multicolumn{3}{c}{Human} & & & \multicolumn{3}{c}{Zero-Shot} & & & \multicolumn{3}{c}{One-Shot} \\
       \midrule
       Adjacent  & & \cOne & \cTwelve & \cEleven & & & \cTwelve & \cOne & \cNine=\cEleven & & & \cTwelve & \cOne & \cEleven \\
       Overlaid & & \cOne & \cThree & \cFour & & & \cOne & \cTwelve & \cEleven & & & \cOne & \cTwelve & \cEleven \\
       Stacked  & & \cOne & \cThree & \cEleven & & & \cEleven & \cOne & \cTen & & & \cOne & \cEleven & \cTwelve \\
       Vertical & & \cOne & \cTwelve & \cEleven & & & \cTwelve & \cOne & \cNine & & & \cTwelve & \cEleven & \cOne \\
       \bottomrule
    \end{tabu}
    \label{tab:top comp types}
\end{table}

\begin{table}[t]
    \caption{Precision@3 for predicting the top comparison types from human takeaways. \add{One-shot precisions are at least as high as zero-shot precisions in all cases.} In the case where two comparison types tied for third place, each was counted as 0.5 comparison types.}
    \centering
    \begin{tabu}{%
  	  c%
  	  	*{2}{r}%
  	}
       \toprule
       Layout & Zero-Shot & One-Shot \\
       \midrule
       Adjacent & 83.33\% & 100.00\% \\
       Overlaid & 33.33\% & 33.33\% \\
       Stacked & 66.67\% & 66.67\% \\
       Vertical & 66.67\% & 100.00\% \\
       \bottomrule
    \end{tabu}
    \label{tab:precision}
\end{table}

\section{Experiment 3: Effect of Context and Data}

In Experiment 2, we calculated the average distributions of LLM comparison types across two examples for each layout, and compared them to distributions derived from human takeaways. The results suggested that LLMs are generally \textit{not} sensitive to spatial layouts like humans are. However, before aggregating the distributions of LLM comparison types, \textbf{we frequently observed significant variations across the two examples within the same layout}. 
For instance, in the stacked layout, while half of the takeaways involved across-group within-element one-to-one comparisons (C1) for one example chart, only one out of 30 takeaways made this type of comparison for the other example (see Figure~\ref{fig:humanllmexample}). 
The two example charts for each layout depicted different contexts (e.g., prices of three drinks in two bars or popularity of three bands in two countries) and different data values. 
% Which of these factors, then, are LLM comparison types sensitive to? 
It is possible that the LLMs overly relied on the context and data to generate takeaways compared to humans. % more than a human would, and thus resulted in inconsistencies between the LLM takeaways and human takeaways. 
% This inconsistency suggests that LLM takeaways depends on the context and data, which could be yet another weakness in LLMs for generating chart takeaways. 
%in LLM comparison type distributions prompted us to pose two questions. 
%Firstly, do humans exhibit the same shifts in comparison types across charts within each layout that we observe in LLMs? If not, this instability could be yet another weakness in LLMs for generating chart takeaways. Secondly, what drives the instability in comparison types across examples in the same layout? 
We ask two questions in this experiment.
First, are humans and LLM similarly sensitive to context and data when generating takeaways?
Second, if not, what drives the instability in comparison types across examples in the same layout?

\subsection{Experiment 3: Procedure, Setup, and Evaluation}

% \subsubsection{LLM instability is not observed in human takeaways}
We focused our analysis on human takeaways and LLM zero-shot takeaways as a case study. To answer the first question, we visualized and compared both human and LLM comparison type distributions for the two examples for each layout. 
For both the human the LLM distributions, we calculated and compared the TVD between the comparison distributions of the two examples. 
We also compared Spearman's rho rank correlation between the frequency rankings of each comparison type across the two examples for both human and LLM distributions.

To answer the second question, we generated eight new bar charts by modifying the original eight stimuli from Xiong~\etal~\cite{xiong2021visual}. For each chart, we preserved its context while introducing a new dataset. For example, if the existing two stacked bar charts visualized ``context~1 and dataset~1'' and ``context~2 and dataset~2'', we created two new stacked bar charts visualizing ``context~1 and dataset~2'' and ``context~2 and dataset~1''. Next, we generated zero-shot takeaways using the optimal configuration established in Experiment~1, and the same coder from Experiment~2 coded these takeaways for comparison types. We then visualized the LLM comparison type distributions for the four example charts for each layout. To examine the effect of contexts on distributions, we computed the TVD's between the distributions for each pair of charts showing the same dataset but different contexts and calculated the average. To examine the effect of data on distributions, we computed the TVD's between the distributions for each pair of charts showing the same context but different datasets and calculated the average.

\subsection{Experiment 3: Results}

We first report on the stability of human comparison type distributions across
examples within the same layout. We then present evidence that variations in LLM comparison type within the same layout is more attributable to context than to data in most cases.

\subsubsection{Human comparison type distributions are stable across examples within the same layout}

Figure~\ref{fig:llmcontextdata} visualizes the human and LLM comparison type distributions for the two example charts from Xiong et al. for each layout. While there is much variation in LLM comparison type distributions across the two examples, human distributions does not appear very different. Table~\ref{tab:tvd2} shows the TVD's between the comparison type distributions for the two example charts in human takeaways and LLM zero-shot takeaways for each layout. In every layout, LLM TVD is larger than human TVD. Table~\ref{tab:rho2} also suggests that while the frequency rankings of human comparison types are all strongly correlated between examples in each layout, those of LLM comparison types are only moderately strongly correlated in two layouts and uncorrelated in the rest. These results confirm that \textbf{the instability in LLM comparison type distributions between examples but within the same layout is not found in human takeaways}. Since the decoding temperature of the GPT-4 was set at 0 when generating these takeaways, this instability was not due to stochasticity in decoding. Hence, we conclude that as far as zero-shot generation is concerned, LLMs are neither accurate nor consistent in writing chart takeaways with human-aligned comparison types. 
\humanllmexample

\begin{table}[t]
    \caption{TVD's between the comparison type distributions for the two example charts in human takeaways and LLM takeaways for each layout. LLM shows much more variation than humans do.}
    \centering
    \begin{tabu}{%
  	  c%
  	  	*{7}{c}%
  	  	*{2}{c}%
  	}
       \toprule
       Layout & Human TVD & LLM TVD\\ 
       \midrule
        Adjacent & 0.2120 & 0.4333 \\ 
        Overlaid & 0.2661 & 0.4333 \\ 
        Stacked & 0.0998  & 0.6333 \\ 
        Vertical & 0.3542 & 0.5000 \\ 
       \bottomrule
    \end{tabu}
    \label{tab:rho2}
    \vspace{-3mm}
\end{table}
% \vspace{-5mm}

\begin{table}[t]
    \caption{Spearman's rho correlation of comparison type rankings in humans and LLM across the two examples for each layout. Significance levels: * p < 0.05, ** p < 0.01, *** p < 0.001. \add{While the rankings of comparison types remain stable across different charts with the same layout in human takeaways, they tend to vary greatly in LLM takeaways.}}
    \centering
    \begin{tabu}{%
  	  c%
  	  	*{7}{c}%
  	  	*{2}{c}%
  	}
       \toprule
       Layout & Spearman's rho (Human) & Spearman's rho (LLM) \\ 
       \midrule
        Adjacent & 0.8937*** & 0.5127 \\ 
        Overlaid & 0.8852*** & 0.7042** \\ 
        Stacked & 0.9285***  & 0.3049 \\ 
        Vertical & 0.9078** & 0.6072* \\ 
       \bottomrule
    \end{tabu}
    \label{tab:tvd2}
\end{table}
% \vspace{-5mm}

\llmcontextdata

\subsubsection{Context affects LLM comparison types more than data}

Figure~\ref{fig:llmcontextdata} shows four distributions of comparison types induced by the four charts for each layout. With the exception of the stacked layout, we see that the distributions visually appear more similar across datasets (same shape, different colors) than across contexts (same color, different shapes). Table~\ref{tab:tvd3} confirms that the average TVD between distributions from different contexts are larger than that between distributions from different datasets with the exception of the stacked layout. Therefore, we conclude that \textbf{LLM comparison type within the same layout is more attributable to context than to data in most cases}.

\begin{table}[t]
    \caption{Average TVD's in LLM comparison type distributions between the two pairs of charts showing the same dataset but different contexts, and the two pairs showing same context but different datasets for each layout. Differing contexts tend to induce greater shifts in LLM comparison type distributions.}
    \centering
    \begin{tabu}{%
  	  c%
  	  	*{7}{c}%
  	  	*{2}{c}%
  	}
       \toprule
       Layout & Different Datasets & Different Contexts\\ 
       \midrule
        Adjacent & 0.2167 & 0.3333 \\ 
        Overlaid & 0.2333 & 0.5000 \\ 
        Stacked & 0.6333 & 0.4667 \\ 
        Vertical & 0.2167  & 0.4833 \\ 
       \bottomrule
    \end{tabu}
    \label{tab:tvd3}
    \vspace{-3mm}
\end{table}

\section{Limitations and Future Work}

To conclude, we detail several limitations from our investigation that lay the foundation for future research.
\vspace{1mm}

\noindent \textbf{Other Chart Types:} \add{In this work, we used bar charts with varying layouts as a case study. While we provide a rich set of human takeaways for each chart, future work can expand to cover more chart types (e.g., scatterplots, line charts, maps) and design manipulations (e.g., color choices, presence of embellishments) to test the generalizability of our results and the capability thresholds of LLM predictions. Since our case study requires much manual coding, we also encourage future work to explore scalable ways to examine LLM perceptual awareness. One promising direction is to employ multiple LLMs to perform qualitative coding and cross-validate each other~\cite{GrundeMcLaughlin2023DesigningLC}. Another direction is to develop automatic pipelines to elicit and model LLMs' visualization-related behavior.}
\vspace{1mm}

\noindent \textbf{More Datasets:} 
\add{The stimuli in the present study were created using two  datasets and eight contexts. To examine the generalizability of our findings, we recommend future research to test a larger number of datasets with additional contexts.
Despite showing poor perceptual alignment on open-domain datasets like the ones we tested, LLMs might produce more human-aligned takeaways on domain-specific datasets. While the pretraining data on open-domain topics include a wide variety of linguistic styles and perspectives, data for certain domains (e.g., protein analysis in scientific visualization~\cite{ho2008interactive}) may feature more formulaic language and stronger patterns. This specificity could make it easier for LLMs to generate takeaways that align closely with human analysis.
Further, we encourage future work to experiment with datasets of varying complexities. Given a more complex dataset, human chart reading behavior tends to diverge as people attend to different aspects of the dataset and visualization, such as locally noticing a statistic or globally noticing a trend~\cite{bearfield2023does, lundgard2021accessible, bearfield2024same}. Thus, employing more complex datasets could similarly offer opportunities to characterize LLM visualization perception on multiple levels and dimensions.}
\vspace{1mm}

\noindent \textbf{Task-Specific Alignment:} 
\add{Existing work has demonstrated that user takeaways with visualizations can be task-dependent~\cite{kim2021towards, malpica2023task}. 
In the present study, we focused on eliciting overall takeaways from charts.
Future work can additionally test whether the misalignment between LLM outputs and human takeaways also takes place when the visualization is provided in the context of a more specific task, such as low-level analytic tasks~\cite{amar2004}. 
Specifically, because humans tend to be prone to cognitive biases even when completing objective analytic tasks with visualizations, such as estimating correlations~\cite{xiong2022seeing, harrison2014ranking}, it would be interesting to explore the extent to which LLMs are capable of predicting these biases in human behaviors.}
\vspace{1mm}

\noindent \textbf{Prompting and Finetuning:} In Experiment 1, we tested for an optimal configuration of chart specifications, temperature, prompting strategies, and input representation using the adjacent bar chart.
Future work can test for optimal configurations across a wider range of charts, layouts, and datasets with an eye towards better generalizability.

Additionally, the two prompting strategies we tested serve as a starting point to demonstrate the power of prompting.
Considering the black-box nature of LLMs, there could be an infinite number of ways to prompt LLMs to obtain even stronger results.
\add{Future work could explore the design space of prompting LLMs for human-aligned chart takeaways.
We suspect adding elements that highlight aspects of the visualization an LLM should pay special attention to, such as pointing a design feature that has been manipulated, would increase LLM-human takeaway alignment. 
Future exploration of the prompt space can also consider manipulating the number of examples provided to the LLMs. 
While the zero-shot and one-shot settings tested in this work are fast and practical, increasing the number of in-context examples has the potential to better calibrate chart takeaways generated by LLMs in light of the benefits of scaling in-context learning in other domains~\cite{agarwal2024many}.
Beyond prompting, another exciting opportunity for takeaway alignment is finetuning. By finetuning LLMs on pairs of charts and human-written takeaways, we might be able to improve their ability to generate human-aligned takeaways.
}
\vspace{1mm}

\noindent \textbf{General-Purpose Visualization Assistants:} \add{Visualization design choices can profoundly impact what people see and take away, making it challenging even for human experts to predict affordances and create effective visualizations~\cite{xiong2021visual}. 
With a deeper understanding of the limitations of LLMs for modeling human perception of visualizations, future work can devise strategies to address these weaknesses and create perceptually-aware, general-purpose visualization assistants. 
Given sufficient knowledge of visualization affordances, LLMs have the potential to generate human-aligned chart takeaways and assist with various other visualization tasks, such as critiquing existing designs, recommending new designs, and annotating visualizations for better comprehension.
Therefore, if we are able to provide appropriate guardrails and develop effective human-LLM collaboration frameworks, LLMs could be able to democratize visualization tasks for laypeople.}

\section{Conclusion}

Through a case study on bar charts with different layouts, we provide an initial attempt to understand the extent to which LLMs are perceptually aware when generating visualization takeaways. In Experiment 1, we identified the optimal configurations to generate chart takeaways. We found that even state-of-the-art LLMs can sometimes struggle to generate semantically diverse and factually accurate takeaways. 
In Experiment 2, we prompted LLMs to generate human-aligned takeaways across multiple spatial layouts. We found that LLMs are generally not sensitive to bar chart layout like humans are.  % when generating takeaways. 
Nonetheless, providing an example chart and some human takeaways helps align model generations in some cases. In Experiment 3, we observed that, across different data and contexts, while humans generally reported similar comparison types across charts with the same layout, LLMs showed tremendous variability in what they compared.
This shows that LLMs are too reliant on context and data when generating takeaways.
We further demonstrated that context affects LLM comparison types more than data for most layouts. 
% \add{add a blurb on limitation and future work}

\acknowledgments{
We thank our reviewers for their helpful feedback. This work was partially supported by NSF award IIS-2237585 and IIS-2311575.
}

\bibliographystyle{abbrv-doi-hyperref}

\bibliography{template}
\end{document}